# An automatic taxonomy of galaxy morphology using unsupervised machine learning

Alex Hocking[1]⋆ James E. Geach[2]† Yi Sun[1] & Neil Davey[1]
[1]*Centre for Computer Science and Informatics Research, University of Hertfordshire, Hatfield, AL10 9AB, UK*
[2]*Centre for Astrophysics Research, School of Physics, Astronomy & Mathematics, University of Hertfordshire, Hatfield, AL10 9AB, UK*

11 September 2017

**ABSTRACT**
We present an unsupervised machine learning technique that automatically segments and labels galaxies in astronomical imaging surveys using only pixel data. Distinct from previous unsupervised machine learning approaches used in astronomy we use no pre-selection or pre-filtering of target galaxy type to identify galaxies that are similar. We demonstrate the technique on the *Hubble Space Telescope (HST)* Frontier Fields. By training the algorithm using galaxies from one field (Abell 2744) and applying the result to another (MACS 0416.1−2403), we show how the algorithm can cleanly separate early and late type galaxies without any form of pre-directed training for what an 'early' or 'late' type galaxy is. We then apply the technique to the *HST* CANDELS fields, creating a catalogue of approximately 60,000 classifications. We show how the automatic classification groups galaxies of similar morphological (and photometric) type, and make the classifications public via a catalogue, a visual catalogue and galaxy similarity search. We compare the CANDELS machine-based classifications to human-classifications from the Galaxy Zoo: CANDELS project. Although there is not a direct mapping between Galaxy Zoo and our hierarchical labelling, we demonstrate a good level of concordance between human and machine classifications. Finally, we show how the technique can be used to identify rarer objects and present lensed galaxy candidates from the CANDELS imaging.

## 1 INTRODUCTION

Machine learning is a data analysis approach that will be vital for the efficient analysis of future astronomical surveys. Even current surveys are generating more data than is practical for humans to exhaustively examine, and the next generation of survey facilities will compound the issue as we usher-in the 'petabyte' regime of astronomical research, with data acquired at a rate of many terabytes per day. For experiments such as the Large Synoptic Survey Telescope (Ivezic et al. 2008, LSST), it will be important to rapidly and automatically analyse streams of imaging data to identify interesting transient phenomena and to mine the imaging data for rare sources will yield new discoveries.

The automatic analysis of galaxies has been a focus of research for some time. Existing, non machine learning approaches to categorise morphology and structure include tools to identify structural parameters such using GALFIT and GIM2D (Peng et al. 2002; Simard 1998) and measures such as the Gini coefficient, M20 and CAS parameters (Abraham et al. 2003; Lotz et al. 2004; Conselice 2003). In addition tools to automate the processing of large datasets to calculate these structural parameters such as Galapagos and Megamorph (Hiemer et al. 2014; Häußler et al. 2013). Other tools use image processing techniques such as Ganalyzer and SpArcFiRe (Shamir 2011; Davis & Hayes 2014). In recent years machine learning has once again become a prominent area of research following high profile advances in object recognition, image classification and generative models (Redmon et al. 2016; He et al. 2016; Goodfellow et al. 2014).

Machine learning has been successfully applied to mundane and complex analysis tasks in astronomy. For example there has been a good deal of effort on developing neural networks and other techniques to improve the estimation of photometric redshifts (Firth et al. 2003; Collister & Lahav 2004; Bonfield et al. 2010; Cavuoti et al. 2012; Brescia et al. 2013). Even the mundane task of automatically classifying objects such as stars and galaxies of different types is well-suited to machine learning as has been recognised for some time, for example, by using neural networks (Lahav et al. 1995) and support vector machines (SVM) galSVM (Huertas-Company et al. 2008; Huertas-Company et al. 2009, 2011). More recent research uses ConvNet models based on work by Krizhevsky et al. (2012) which won the computer science ImageNet competi-tion in 2012. This has been used to classify SDSS images (Dieleman et al. 2015, Galaxy Zoo) and Huertas-Company et al. (2015) followed Dieleman's work to classify CANDELS galaxies into five morphological types. Convolutional networks have been used more recently to classify radio galaxies (Aniyan & Thorat 2017). Other machine learning algorithms such as deep recurrent networks have be used to classify supernovae (Charnock & Moss 2017). Random forests have been used to identify transient features in Pan-STARRS imaging (Wright et al. 2015), for the identification and classification of Galactic filamentary structures (Riccio et al. 2016) and the inference of stellar parameters (Miller et al. 2015). Kuminski & Shamir (2016) created a catalogue using a tool called Wndcharm to classify ∼3,000,000 SDSS galaxies as spiral or elliptical. More recently Generative Adversarial Networks (GANs) (Goodfellow et al. 2014) have been employed by Schawinski et al. (2017) to de-noise images of galaxies with much greater performance than simple deconvolution.

⋆ a.hocking3@herts.ac.uk
† j.geach@herts.ac.uk

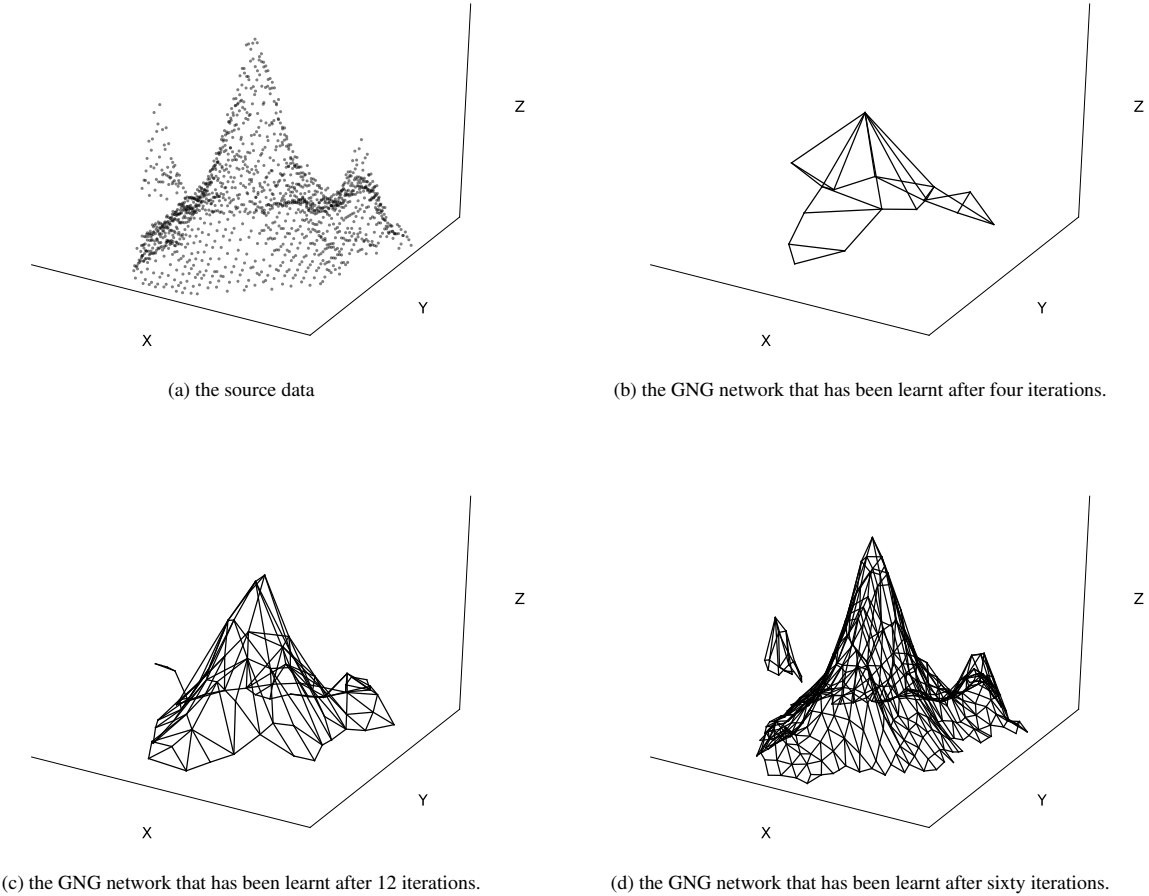

**Figure 1.** These four images show how the Growing Neural Gas (GNG) algorithm works to map and approximate data. (a) shows the sample data. The images (b), (c) and (d) show the progress of the GNG algorithm as it discovers and learns the structure of the data.

These techniques predominantly employ *supervised* learning. Supervised learning has the disadvantage that it requires labelled input data, and so is limited in its potential for completely automated data analysis and exploration of large data sets. For example, Huertas-Company et al. (2015) created a CANDELS catalogue which relied on 8000 expert classifications of galaxies in GOODS-South to train the ConvNet to classify the remainder of the CANDELS into five visual-like morphologies: disk, spheroid, peculiar/irregular, point source/compact and unclassifiable. These upfront classifications drive the process and a ConvNet cannot classify objects outside of these pre-defined labels used in the training process. Unsupervised machine learning offers an alternative approach. It enables exploratory data analysis eliminating the need for human intervention (e.g. pre-labelling). The potential for this has been recognised for over two decades (Klusch & Napiwotzki 1993; Nielsen & Odewahn 1994; Odewahn 1995).

Unsupervised learning has already found application in astronomy, particularly in the estimation of photometric redshifts (Geach 2012; Way & Klose 2012; Carrasco Kind & Brunner 2014), object classification from photometry or spectroscopy (D'Abrusco et al. 2012; in der Au et al. 2012; Fustes et al. 2013), finding galaxy clusters using catalog data (Ascaso et al. 2012) and searching for outliers in SDSS galaxy spectra (Baron & Poznanski 2016). Work by Schutter & Shamir (2015) presents computer vision techniques to identify galaxy types (see also Banerji et al. 2010). This approach required an existing catalogue of galaxy images that are sorted by class at the input stage, which is pre-labelling and therefore a super-vised process. Other work by Shamir (2012) developed an outlier detection technique to detect peculiar galaxies among a training set consisting of a single clean morphological type. The technique trains unsupervised algorithms on a pre-labelled and collated training set. Shamir et al. (2013) used a pre-defined training set with supervised and unsupervised algorithms to classify galaxy mergers. (Shamir & Wallin 2014) combined supervised and unsupervised techniques in an outlier technique to identify peculiar galaxy pairs in 400,000 SDSS images. Existing work incorporating unsupervised algorithms to classify images of galaxies all use the collation of a training dataset by pre-labelling galaxies. A completely unsupervised machine learning technique that can be applied to survey images without this upfront effort is arguably yet to be proven.

In this paper we employ a patch based unsupervised machine learning model to explore surveys by classifying, labelling and identifying similar galaxies. The technique reads multi-band FITS survey images and outputs groups of similar galaxies. The technique combines small patches around each pixel where each small patch is typically much smaller than the size of a galaxy. To the best of our knowledge no other unsupervised, or supervised machine learning technique has used a patch based model. Unlike previous unsupervised machine learning approaches our technique is an exploration and classification approach to produce catalogues consisting of fine-grained classifications of whole surveys. The technique is completely unsupervised requiring no up-front collation of training data and no galaxy pre-labelling.

We demonstrate the technique by applying the algorithm to *Hubble Space Telescope (HST)* Frontier Fields (FF)[1] observations of two massive clusters of galaxies. These are fields that contain a mixture of early and late type galaxies that offer an ideal test case and we use this data to demonstrate the principles of the method. We then apply the technique to the five *HST* CANDELS fields, producing a hierarchical classification catalogue for approximately 60,000 sources. We note that he *HST* CANDELS fields have been automatically classified before. Van der Wel et al. (2012) provided structural parameters for galaxies in CANDELS using the GALAPAGOS software (Barden et al. 2012) and the aforementioned CANDELS catalog produced by Huertas-Company et al. (2015) using supervised machine learning that consists of five classification types. Our catalogue is distinct from the catalogue of (Huertas-Company et al. 2015) because it provides a finer classification of galaxies grouped by morphology and photometric characteristics. The unsupervised machine learning technique also enables a galaxy similarity search that to our knowledge has not been demonstrated before using unsupervised machine learning techniques. The resulting catalogue, visual catalogue and galaxy similarity search is provided at www.galaxyml.uk and the source code at https://github.com/alexhock/galaxymorphology

Finally, we prove how the technique is not only useful for cataloguing survey images but also for identifying rarer objects by revealing two lensed galaxy candidates (See §4.1). To our knowledge these galaxies have not been previously identified as lenses.

The paper is organised as follows: in §2 we describe the algorithms in more detail, in §3 we describe the process of applying the algorithms to automatically identify early and late type galaxies in the FF, and in §4 we present the analysis of applying the technique to the CANDELS fields, the catalogue and a comparison with the Galaxy Zoo: CANDELS project classifications. We conclude in §5 with a comment on the limitations of our method and avenues for future development.

## 2 THE ALGORITHMS

In this section we introduce the three algorithms that comprise the overall method. There are two unsupervised machine learning algorithms and one image processing algorithm. The input and output of the machine learning algorithms are compatible and therefore the algorithms can be chained together in multiple configurations where, for example, the output of one algorithm can be used as the input to another. We describe an application of the method that does this in §3.

### 2.1 Growing Neural Gas

The Growing Neural Gas (GNG) algorithm (Fritzke et al. 1995) creates a graph that represents the latent structure within data. The algorithm is used for the purposes of clustering and analysis of many types of data. GNG is applied to an $m \times n$ data matrix representing the input data that contains $m$, $n$-dimensional vectors called sample vectors. GNG identifies structure by iteratively growing a graph to map the data in the sample vector space. The graph consists of nodes connected by lines called edges. Each node has a position in the data space called a code vector. This is illustrated in Figure 1. The code vectors have the same dimensionality as the sample vectors in the data matrix. The algorithm starts by creating a graph of two nodes. Each node is initialised using a random sample from the data matrix. The graph grows and shrinks as the input data is processed (i.e. more samples are introduced). During this process the positions of the nodes evolve: the code vectors are updated to map the topology of the data and the graph splits to form disconnected sub graphs, each of which represents a cluster in the data space. The process continues until a stopping criterion has been met, such as a saturation value for the number of nodes within the graphs, or the processing time. In order to create a graph that accurately maps the input data it is common to process the input data multiple times. The learning steps of the algorithm are:

(i) *Initialization* Create a graph with two nodes. Initialise the position of each node with the vector of values from a random sample vector **p** from the data matrix. Subsequently, samples are drawn at random from the data matrix and the following set of rules applied:

(ii) *Identify the two nodes nearest to the sample vector* For each node in the graph, the distance $d$ between the sample vector **p** and the node's code vector **q** is calculated using the squared Euclidean distance. The two nodes $(s_0, s_1)$ most similar to the sample vector (i.e. the two smallest values of $d$) are identified.

(iii) *Create and update edges* If an edge connecting $s_0$ and $s_1$ does not exist, create it. Set the 'age' of the edge connecting $s_0$ and $s_1$ to zero. Increment the age of all other edges connected to the nearest node $s_0$.

(iv) *Increase the 'error' of the nearest node $s_0$* The 'error' is simply the squared Euclidean distance between a sample vector and nodes in the GNG: if the error is high then the GNG has not yet properly mapped the data space containing the sample vector. In this step the squared Euclidean distance between the input sample vector and $s_0$ is added to the local error of $s_0$.

(v) *Move the nearest node $s_0$* Update the code vector of $s_0$ using equation 1. This step moves the nearest node $s_0$ 'towards' the

---

[1] https://archive.stsci.edu/prepds/frontier/

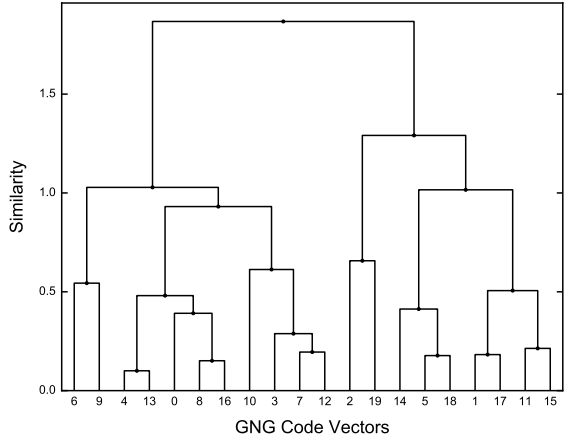

**Figure 2.** A dendogram visualisation of the hierarchy identified by the hierarchical clustering process. The $x$ axis gives the node identifiers from GNG. The $y$ axis represents the degree of similarity. The root node is shown at the top.

input sample vector **p**. The $\epsilon_b$ parameter controls the size of the movement towards the input sample.

$$\Delta \mathbf{q}_{s_0} = \epsilon_b (\mathbf{p} - \mathbf{q}_{s_0}) \qquad (1)$$

(vi) *Move connecting nodes' neighbours* Using the same process as in the previous step but using the $\epsilon_n$ parameter to control the magnitude of the adjustment for nodes directly connected by an edge to $s_0$.

(vii) *Remove old edges and nodes* Remove all edges with an age greater than the maximum age $A$ parameter. All nodes without edges are removed.

(viii) *Add a new node to the GNG graph* A new node is added to the graph after a fixed number ($\lambda$) of sample vectors have been processed. The new node is added at the midpoint between the node with the highest error and its connecting node. If multiple nodes are connected then the new node is positioned at the midpoint of the connecting nodes with the highest error. When a new node is added, the error of each node is reduced by $\alpha$.

(ix) *Reduce all node error values* Reduce the error of each node in the GNG graph by a factor of $\beta$.

Fritzke et al. (1995) describes the parameters mentioned above in detail. The majority of the compute time is in step (ii); various attempts have been made to reduce the time taken (Fiser et al. 2012; Mendes et al. 2013). For a data matrix with few dimensions using various tree based methods to store GNG nodes works well and provides a significant performance increase over a brute force method to identify the nearest neighbours in the first step of the algorithm. However, as the dimensionality of the data increases the performance of the graph based methods decreases to become similar to that of the brute force method. We implemented a version of GNG that parallelises the brute force method of finding nearest neighbours as this provides the most flexibility.

### 2.2 Hierarchical Clustering

Hierarchical clustering (HC) (Hastie et al. 2009) involves a recursive process to form a hierarchical representation of a data set (e.g. the code vectors of the nodes output by GNG) as a tree of clusters.

One of the key benefits of HC is that it can produce uneven clusters, both in terms of their disparate sizes and separation in the parameter volume. Many unsupervised learning algorithms produce even cluster sizes which imply an assumption about the structure of the data; HC makes no such assumption. The identified clusters form a hierarchical representation of the input data, as illustrated in Figure 2. This hierarchical representation can be thought of as a tree structure where the leaves represent the individual input sample vectors from the data set. The process starts by merging pairs of leaves, using a measure of similarity to identify the most similar pair of leaves. The pair with the closest proximity are merged into a new cluster (twig) that is added to the tree as a new parent node to the pair. The process continues by merging pairs of nodes at each level until a single node remains at the root of the tree. The final tree representation contains multiple 'levels' of nodes, with each node in a level representing a cluster. Each level can be considered a level of detail in a clustered representation of the data. Our approach is to apply HC to the *output* of the GNG, further refining this representation of the input data into a cluster hierarchy that can be used to segment and classify image components.

There are a number of methods used to measure similarity between vectors, including Euclidean distance, Pearson correlation and cosine distance. After experimenting with these three types we found the best results were obtained using the Pearson correlation coefficient (see equation 2) and cosine similarity (see equation 3) measures,

$$r(\mathbf{p}, \mathbf{q}) = \mathrm{cov}(\mathbf{p}, \mathbf{q}) \mathrm{var}(\mathbf{p})^{-0.5} \mathrm{var}(\mathbf{q})^{-0.5} \qquad (2)$$

where $r$ is the Pearson correlation between **p** and **q** (the code vectors from two GNG graph nodes) and

$$\cos(\theta) = \frac{\mathbf{p} \bullet \mathbf{q}}{\|\mathbf{p}\|\|\mathbf{q}\|} \qquad (3)$$

the cosine similarity is the cosine of the angle between the two vectors. The process of using a similarity measure to merge clusters is called linkage. We apply 'centroid' linkage which uses the chosen similarity measure to compare the centroids of the clusters at each level of the tree; a centroid is calculated by finding the average sample value within a cluster. After assessing the pairwise distance between all clusters in a level, clusters with the minimum linkage are merged, and the centroid of the merged cluster recalculated, ready for the next merging step as one moves up the hierarchy towards the single root.

Each node in the tree can be given a unique label and so the input data can be classified according to which node in the tree best describes it, at some desired 'level of detail' (the trivial example is that the 'root' by definition would label all of the data). In this work we are concerned with imaging data, and the algorithm described above can be used to label individual (or groups) of pixels in an image, therefore automatically segmenting and classifying them. Consider an image containing two different types of object: provided the data matrix captures the difference between these objects (be it morphology, colour, intensity, etc.), then the algorithm described above should automatically identify and label these two objects differently.

### 2.3 Connected-component labelling

Connected-component labelling is a general term used to describe a process that can identify and label sub-structures within a data set. Each sub structure is called a component and consists of a set of connected data elements which are considered to be connected if they are joined in some way (for example, vertices that are connected by an edge in an undirected graph). A typical result of the process is a list of uniquely labelled components each consisting of a sub-set of the data elements, where no data element is shared by more than one component. The algorithm is commonly used in image processing to identify and label connected groups of pixels, for example, to identify and extract blobs in binary images. It's not clear when the connected-component labelling concept originated but it has been in use since the 1970s, for example in Hoshen & Kopelman (1976).

Although the general concept is fairly straightforward there are a surprising number of implementation options. Much work has been carried out such as a) the efficient tracing of component outlines or contours (Chang et al. 2004) and b) investigations into algorithm efficiency, considering the relative merits of using a single pass, two pass, or even multiple passes through the data elements (He et al. 2008; Wu et al. 2009). One would expect a single pass algorithm to be the most efficient, however due to the non-sequential memory accesses required by the single pass algorithm the two pass algorithms remain very competitive and execution-time scales linearly with the number of data elements (Wu et al. 2009). Other areas of research into algorithm efficiency include identifying efficient data structures to store and attach labels to data elements such as the 'union-find' data structure (Fiorio & Gustedt 1996) and research into the parallelisation of various connected-component algorithms including the use of GPUs using NVIDIA's CUDA (Kalentev et al. 2011).

We implement an efficient, sequential version of the algorithm inspired by parts of Wu et al. (2009); Fiorio & Gustedt (1996). However, we deviate from the standard implementations used in image processing by using the algorithm on sub-images (thumbnails) instead of pixel data. Therefore, the term 'data element' in the previous and following sections refers to an individual sub-image. The algorithm proceeds by iterating through the data elements and assigning a label, consisting of an integer value, to each data element. The following steps are performed for each element (the first pass):

(i) Retrieve the labels of the neighbouring data elements. Any overlapping or adjacent data elements are neighbours. Overlapping and adjacent data elements can be identified using their positions and size.

(ii) If there there are no neighbours or none of the neighbouring data elements have labels then create a new label with a unique identifier (an integer that starts with a value zero, incremented for each new label) and apply it to the data element. Continue to the next data element.

(iii) If any neighbouring data elements have labels then identify the neighbouring label with the smallest unique identifier and assign the label to the data element.

(iv) Add the unique labels of the neighbouring elements to a list called an equivalence list. This list is used at the end of the process to identify all the labels that belong to the same component.

At this point every data element has a label (which may also be shared among many other data elements), each label belongs to an equivalence list and each equivalence list contains all the labels for a unique component. The second pass is purely a re-labelling process to ensure that every data element in a component has the same label. It proceeds by identifying the equivalent list that the label of each data element belongs to, finds the label in the list with the smallest identifier (the find function of the union-find data structure) and then applies that label to the data element. The output of the algorithm is a list of components and their data elements. The location and size of all the data elements are known and therefore these lists can be used to identify the properties of a component, for example, the width, height and an approximation of its centre.

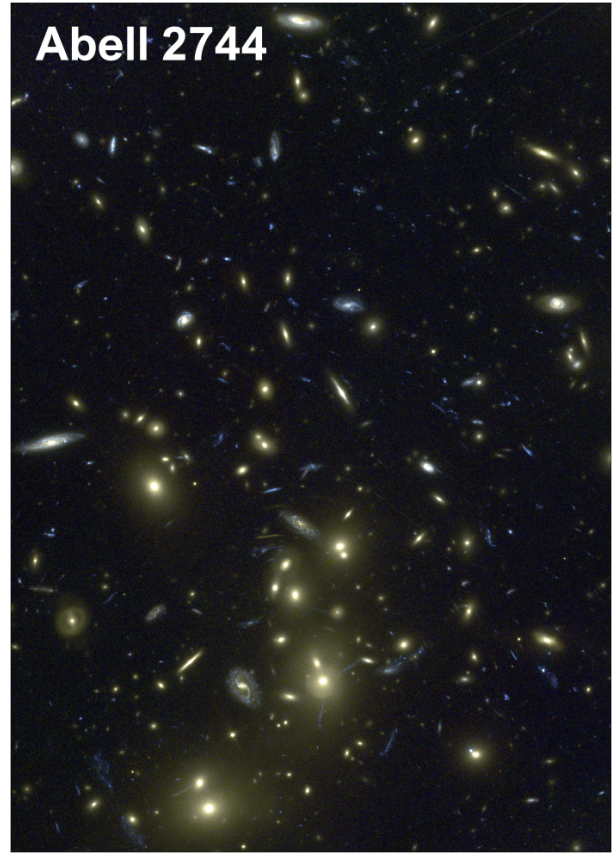

**Figure 3.** Training data for our demonstration example. This is an RGB composite image of the *HST* Frontier Field Abell 2744 ($90'' \times 130''$). The red, green and blue channels correspond to the F814W, F606W and F435W bands. We chose this data set as it represents a classic example of object segregation that is well understood: the cluster dominated by red elliptical galaxies, surrounded by blue late types and gravitationally lensed features. In our proof-of-concept the goal is to demonstrate that the algorithm can cleanly classify these two basic classes automatically in much the same way a human inspector would. Importantly, since the Frontier Fields target several clusters, we can test the algorithm on a different, 'unseen' cluster.

## 3 APPLICATION: THE HUBBLE FRONTIER FIELDS

### 3.1 The data

We use deep *Hubble Space Telescope (HST)* images (F435W, F606W and F814W bands) of the strong lensing galaxy clusters Abell 2744 and MACS 0416.1-2403 to demonstrate a proof-of-concept and practical application of the algorithm. Since images of clusters contain two distinct galaxy populations (namely bright, red early types and numerous blue late types comprised of cluster members and background galaxies, including gravitationally lensed features), these data provides an excellent opportunity to test whether the algorithm can automatically identify and dis-tinguish these classes of object that a human could trivially do 'by eye'.

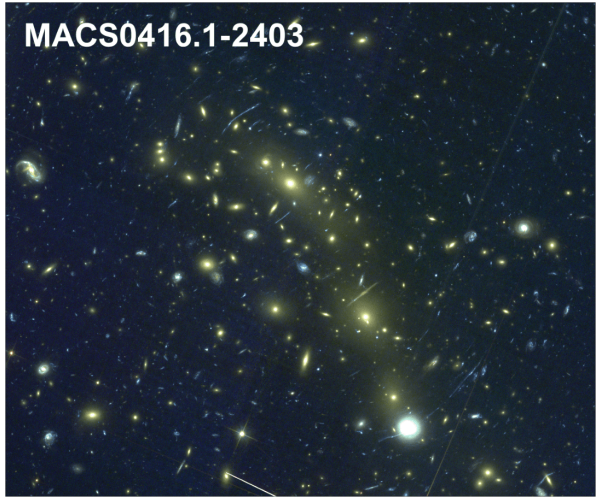

**Figure 4.** An RGB (F814W, F606W, F435W) composite image of the Frontier Field MACS 0416.1-2403 ($160'' \times 130''$). We have applied the algorithm to this new 'unseen' data after training on Abell 2744 (Fig 3) to automatically classify two groups of sources that correspond to classical early and late types (Fig 5 & 6).

### 3.2 An overview of the process

As described in §2 unsupervised learning algorithms such as GNG and HC can be used to identify latent structure within data. The output of a successful unsupervised learning process is merely a condensed version of the original data set that retains the majority of the original data's structure. Therefore we can use these algorithms to take a large data set and effectively reduce it to a size that can be processed more efficiently. Additionally, these learning algorithms can find areas of high density within the data space, and since GNG is graph based, they can also determine whether these dense areas are spatially separated. Regions of high density are called clusters and represent a subset or grouping of the original data set. They can also be considered to be a subset of latent structure within the data. For simple data analysis the identification of clusters may be all that is required, but we can go further and also consider whether the clusters can be used to identify similar structure in new unseen data. For example, when analysing astronomical images such as the FFs, we can identify features that exists in one image (encoded into the output of the GNG+HC), and then use that information to identify similar galaxies in a new unseen image.

#### 3.2.1 The learning phase

The goal of the learning phase is to automatically discover the different groups of galaxies that exist in a survey image containing hundreds of galaxies by using pixel data alone. The discovered groups can then be used to analyse new survey images to identify the same types of galaxies.

An important aspect of the design of the learning phase is how to represent the data from the training survey images at each step. We can use the source data directly or use more complex feature extraction processes that emphasise the characteristics of the data we need the system to learn. Our goal is that the algorithm automatically learns the types of galaxies that exist in the training image. Therefore, we construct the learning phase to exclude information about the angular size and orientation of galaxies so that the system groups galaxies into types using other general characteristics such as colour and morphology. The learning phase proceeds as follows:

(i) *Convert the entire survey image into a data matrix*  From a survey image containing possibly 100s of galaxies create an $m \times n$ data matrix (DM1) where each row, an $n$-dimensional sample vector, is a rotationally invariant representation of a small sub image patch, typically much smaller than a galaxy. A sample vector is created for a patch around each pixel of a galaxy. Therefore there is a dense, oversampling of sub image patches. Each column in the data matrix is known as a feature. Features that have a much larger range of values than others will dominate the distance calculations used in GNG and Hierarchical Clustering. Therefore, the features of the data matrix are normalised to have zero mean and unit standard deviation.

(ii) *Apply GNG to the data matrix*  GNG creates an accurate topological map of data in matrix 1 (DM1). The output consists of another $k \times n$ data matrix (DM2), where $k$ is the number of GNG nodes and $k < m$. Each sample vector in DM2 represents a cluster (group) of similar small sub image patches used to create DM1.

(iii) *Apply Hierarchical Clustering to the output of GNG*  Further reduce the number of groups by using HC to identify the groups within DM2. Each identified group represents a sub-set of the samples vectors of DM2. Therefore, each grouping can be thought of as a 'type' of sub image patch.

(iv) *Apply connected-component labelling to identify galaxies*  Identify the numbers and types of the small sub-image patches that form each galaxy in the survey image. Many small sub-image patches form each galaxy. This step can be performed after DM1 has been created.

(v) *Create a galaxy data matrix*  Create a new data matrix (DM3) by creating a sample vector for each galaxy. Each element in a galaxy's sample vector corresponds to one 'type' of sub-image patch. The value of each element is the number of small sub-images in the galaxy of that 'type'. Many sub images patches form each galaxy (one for each pixel of the galaxy) and each sample vector is a histogram of the types of sub images contained in a galaxy. Each of the sample vectors is normalised to improve scale invariance by dividing each element by sum of the sample vector. The result is a data matrix where each sample vector is a scale and rotation invariant representation of a galaxy. The rotation invariance occurs because the relative positions of each patch that forms the galaxy is not included in the galaxy's histogram representation.

(vi) *Identify the groups of galaxies*  Hierarchical clustering is then used a second time to identify the main groups (or types) of galaxies that exist in DM3. The position of each group in data space is recorded and used in step (iv) of §3.2.2 to identify the type of galaxies in new images.

#### 3.2.2 Identifying galaxies in new 'unseen' images

At this point the input training images have been processed and the types of galaxies automatically identified using pixel data alone. We can now locate and classify large numbers of galaxies in new 'unseen' survey images by using the new survey image pixel data and the information obtained in the learning phase. This process is summarised as follows:

(i) *Convert the unseen image into a data matrix*  Convert the

new image into a new data matrix (DM4) using the same process used to create DM1 in step (i) above.

(ii) *Classify the sub-image patches*  Compare the DM4 sample vectors with the DM2 sample vectors by using an efficient nearest neighbour search. Each DM4 sample vector will assume the HC 'type' of the most similar sample vector in DM2.

(iii) *Apply connected-component labelling to identify galaxies*  Use the connected-component labelling and data matrix generation process described in step (iv) of the learning phase to create a data matrix (DM5) that contains a sample vector for every galaxy in the image. In addition, output a catalogue of the galaxies which includes their approximate dimensions in pixel space.

(iv) *Classify each galaxy*  Each galaxy sample vector in DM5 assumes the type of the most similar galaxy type identified in step (vi) of the learning phase.

The following section describes the complete process applied to the FF.

### 3.3 The Learning Phase Applied to Frontier Field Abell 2744

#### 3.3.1 Pre-processing

The input data matrix (DM1) consists of sample vectors that comprise a sequence of $8 \times 8$ pixel thumbnails sampled from each of the training images (the aligned F435W, F606W and F814W images of Abell 2744, Figure 3). Tests on various sizes of thumbnails found that eight pixel square thumbnails produced the best results in terms of processing speed and galaxy detection; using larger patches resulted in a reduction in the identification of very small galaxies (effectively, this is a resolution issue). For each thumbnail we evaluate the radially averaged power spectrum of pixel values in five bins, allowing us to encode information about the pixel intensity in a manner that is rotationally invariant. The power spectrum for each filter is concatenated into a single 15-element sample vector, that naturally encodes colour information to the data matrix. Thus the data matrix consists of rows of sample vectors and 15 columns called feature vectors.

To improve speed, during training we only consider regions of the image with pixel values in excess of 5 times the root mean squared value of blank sky in the image[2]. This reduced the number of image thumbnails to 851,000. Note that these thumbnails consist of small sections of galaxies and not whole galaxy images. Histograms of the feature vectors displayed log normal distributions. In order to convert each feature to a normal distribution, thus creating a better clustering outcome, we simply took the natural log of values in the data matrix. Each of the feature vectors were then normalised by subtracting the mean and dividing by the unit of standard deviation.

#### 3.3.2 GNG & Hierarchical Clustering

We configured the maximum nodes parameter of the GNG algorithm (§2.2) to 20,000 and processed each of the 851,000 sample vectors 100 times. The output of this step is data matrix (DM2) of $20,000 \times 15$, representing the code vectors of the GNG nodes. DM2 is then used as input into the HC algorithm (§2.2). The HC was run with three types of similarity measure including: Euclidean distance metric, cosine similarity measure and the Pearson correlation coefficient, with the Pearson correlation coefficient (see equation 2) achieving the best results. We searched down the resulting hierarchical tree from the root node to identify the relevant child groupings (clusters) of GNG nodes. Each cluster contained a corresponding error value which indicated the 'quality' of the cluster. We selected all the clusters that had an error of 0.15 or less, which identified 536 independent clusters of GNG nodes. Using a higher error value would identify fewer clusters that contained larger numbers of GNG nodes. However, the next steps in the process are not sensitive to larger numbers of clusters and therefore we chose a smaller error value which represented higher quality clusters that are more accurate (i.e. the GNG nodes are more similar). Using GNG and HC we have identified 536 groups that contain the original population of 851,000 sub-images.

#### 3.3.3 Connected-component labelling

We used the connected-component labelling algorithm described in (§2.3) to identify spatially connected sub images (components) in DM1. These connected sub-images represent the individual galaxies. The FF images contain crowded central fields with bright, extended stellar halos around elliptical galaxies. In order to separate the galaxies in the central elliptical cluster we identified two thresholded lists of the 851,000 sub-images. One list identified the sub images at locations with pixel intensity of at least $5\sigma$ over the background level and a second list identified the image patches at least $10\sigma$ over the background level (where $1\sigma$ is the root mean square value of the source-free background). The locations of the 851,000 sub-images were identified and the mean pixel intensities from each of the three bands were compared to the threshold level. If any of the pixels were over the threshold level in any of the three bands the sub-image patch was added to the list.

The connected-component labelling process used the following inputs i) the co-ordinates of each of the 851,000 sub-image patches ii) the size of the sub-image patches ($8 \times 8$ pixels) iii) a minimum component size of five, so that only components with five or more sub-images were considered, and iv) the $5\sigma$ and $10\sigma$ threshold lists. Any component overlaps were identified and the $10\sigma$ component was selected in preference to any overlapping $5\sigma$ component. This enabled the galaxies in the brightest areas of the extended stellar halo to be distinguished. A catalogue of the components was created by calculating the approximate position of the component (calculated using the average position of its sub images) and the width and height of the component was calculated by identifying the minimum and maximum co-ordinates of the sub images.

#### 3.3.4 Identifying galaxies

The next step combined the 536 clusters of sub images from the HC process and the components identified by the connected component labelling process to create a new data matrix (DM3). Each sample vector in the data matrix represented a component (galaxy) consisting of 536 elements. The value of each element was a count of the number of sub-images in the component that was in the representative hierarchical cluster. The resulting sample vectors were sparse in that the majority of the elements were zero. The final preparation step used to create DM3 was a normalisation: divide each element in the sample vector by the sum of all its elements. A large galaxy and small galaxy of the same type will consist of the same types of sub-images (identified by HC). However, there will be a large difference in sub-image counts in each element. Therefore we divided each element in a sample vector by the sum of the vectors elements which rescaled the vector elements to de-emphasise galaxy size.

---

[2] Although note that in principle this data could be used during training

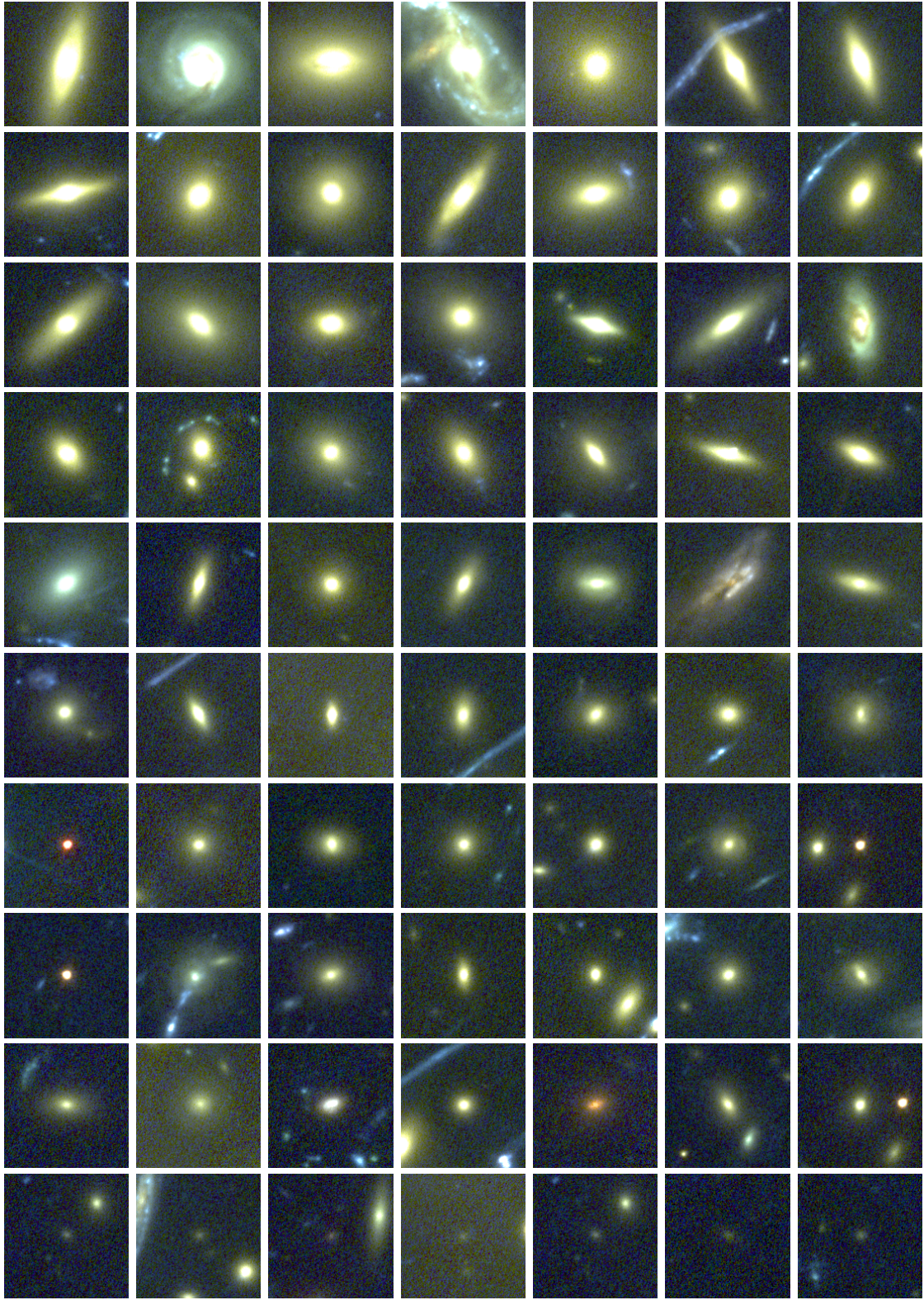

**Figure 5.** Examples of a sample of galaxies in MACS0416.1−2403 that the algorithm automatically identifies as being members of group 'one'. Each image is $4.5'' \times 4.5''$. The algorithm automatically identified this group and classified these galaxies using no data other than the image pixel intensity values from the F435W, F606W and F814W bands, and based classifications on the information in the Abell 2744 image.

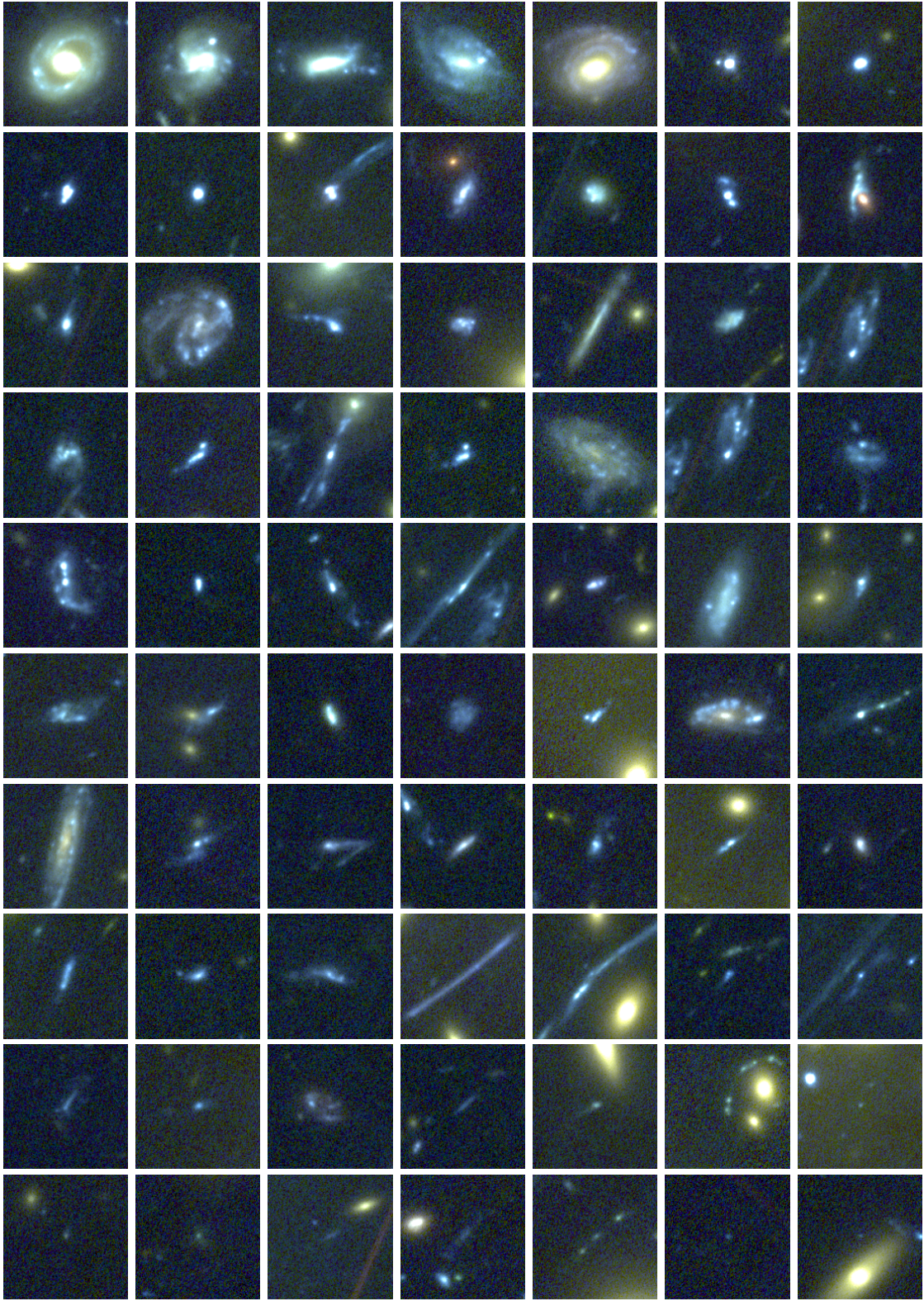

**Figure 6.** Examples of a sample of galaxies in MACS0416.1−2403 that the algorithm identifies as being members of group 'two'. Each image measures $4.5'' \times 4.5''$ arcseconds. Lensed galaxies are included in this group. Again, the algorithm automatically identified this group and classified these galaxies using no data other than the image pixel intensity values from the F435W, F606W and F814W bands, and based classifications on the information in the Abell 2744 image. Note that in some cases the algorithm has correctly classified faint galaxies that are clearly in the stellar halo of an elliptical.

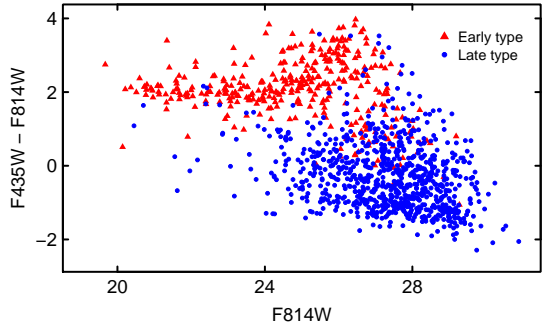

**Figure 7.** A colour-magnitude diagram of the galaxies in MACS 0416.1−2403. The galaxies that the process identifies as being members of group 'one' are labelled with the red triangles. The galaxies that the process identifies as members of group 'two' are labelled with blue circles. The process cleanly separates the early types in the red sequence and the late types in the blue cloud.

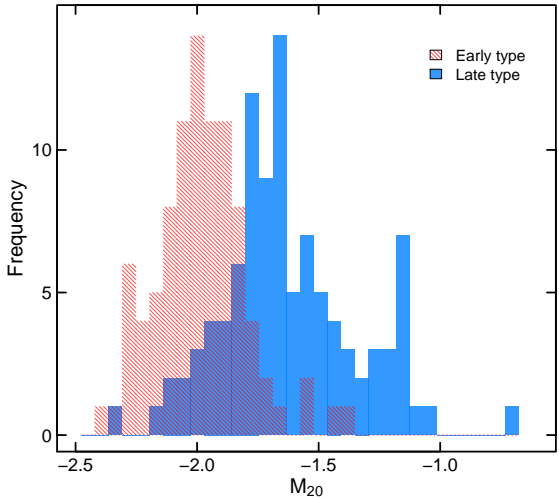

**Figure 8.** Histograms showing the $M_{20}$ morphological measure calculated for the galaxies that the process identifies as being members of group one in red, and the galaxies that the process identifies as being in group two in blue. This appears to identify two populations of galaxies as found in Lotz et al. (2004).

The final step was to use HC again on DM3 to identify 'clusters' of galaxies that are similar to each other, using the cosine similarity measure. Cosine similarity is a measure of the angle and not magnitude between two vectors and therefore improves the scale invariance of the process. We ran the algorithm with a pre-set parameter to output the two best clusters or groups of galaxies.

### 3.4 Verifying the method

The learning phase identifies two groups of galaxies in the Abell 2744 images broadly representing late type (blue spiral, irregular, lensed) and early type (red, smooth, elliptical) galaxies. We then used the trained network to analyse a new, unseen image of the same type (MACS 0416.1−2403) by performing the steps in §3.2.2. This analysis identified the same two groups of galaxies and produced a catalogue of the galaxies and their type. Example galaxies from these two groups are shown in Figures 5 and 6. No pre-existing labels are available, therefore typical measures used in supervised machine learning to analyse accuracy such as precision/recall are not available. Instead, in order to verify the results, we investigate how the method compares to two traditional techniques for classifying early/late type galaxies. First, the two classes of galaxy should be cleanly separated in a colour-magnitude diagram, and indeed we find this is the case (Fig 7.). Photometry was measured using SExtractor on cut-outs of each galaxy in the classified sample. The figure shows the algorithm correctly identifies the red sequence and blue cloud, although clearly with some scatter between the point clouds; generally these are due to close blends and projections. We also calculated the $M_{20}$ morphological parameter Lotz et al. (2004) for galaxies larger than $15 \times 15$ pixels in the F814W band. $M_{20}$ is the normalized second order moment of the brightest 20% of the source flux, with less negative values corresponding to clumpier sources. Figure 8 shows the results, which shows a systematically lower $M_{20}$ value for our early types compared to our late types. We argue that Figures 5–8 demonstrate the proof-of-concept success of the algorithm in automatically classifying sources into astrophysically meaningful groups. In the following we apply this method to a broader input set – the *HST* CANDELS fields.

## 4 CLASSIFYING THE CANDELS FIELDS

The Cosmic Assembly Near-infrared Deep Extragalactic Legacy Survey (CANDELS) (Grogin et al. 2011; Koekemoer et al. 2011) is an *HST* survey designed to document the evolution of galaxies out to $z \approx 8$. The survey consists of Wide Field Camera 3 optical and infrared (WFC3/UVIS/IR) and Advanced Camera for Surveys (ACS) optical imaging of five extragalactic survey fields. There are two tiers: a 'deep' survey to at least four orbit effective depth in F160W over $\sim$125 arcmin$^2$ in GOODS-South and GOODS-North, and a wider shallower survey to two orbit effective depth covering $\sim$800 arcmin$^2$ of COSMOS, EGS and UKIDSS/UDS and flanking areas of GOODS-South. For all five fields we used version 1.0 release of the data[3], selecting the filters F160W and F814W as they provide the most complete coverage across all five fields. The F814W images were projected onto the same grid as the F160W (0.06″ per pixel).

The process to analyse the FF images (§3) used generalisation by training on one field and then then applying the model to classify objects in a second field. We took this approach as it was important to prove that it is possible to do this using an unsupervised approach in order to significantly reduce processing time as the computational time for applying GNG and HC on very large data would be prohibitive. However, when considering the size of the CANDELS dataset (F160W and F814W imaging) we see that it is fairly small at $\sim$60 Gb and therefore we apply the learning algorithms to all five fields of the CANDELS data in its entirety.

Before describing the CANDELS classification process we point out that combining the data from the deep and wide fields is not ideal for machine learning processes. The initial assumption is that data is prepared in a consistent manner. In this case, the depth of the images varies across the fields and in some cases the classification process identifies groups that contain galaxies predominantly from GOODS-North and GOODS-South and other groups predominantly from UDS, EGS and COSMOS. In §4.1 we compare our catalogue to the Galaxy Zoo: CANDELS classifications and we note that the Galaxy Zoo team has provided alternative weighted classifications for the galaxies in the deep sections of the survey, illustrating that the combination of depths appears to affect human classifiers too.

---

[3] https://archive.stsci.edu/prepds/candels/

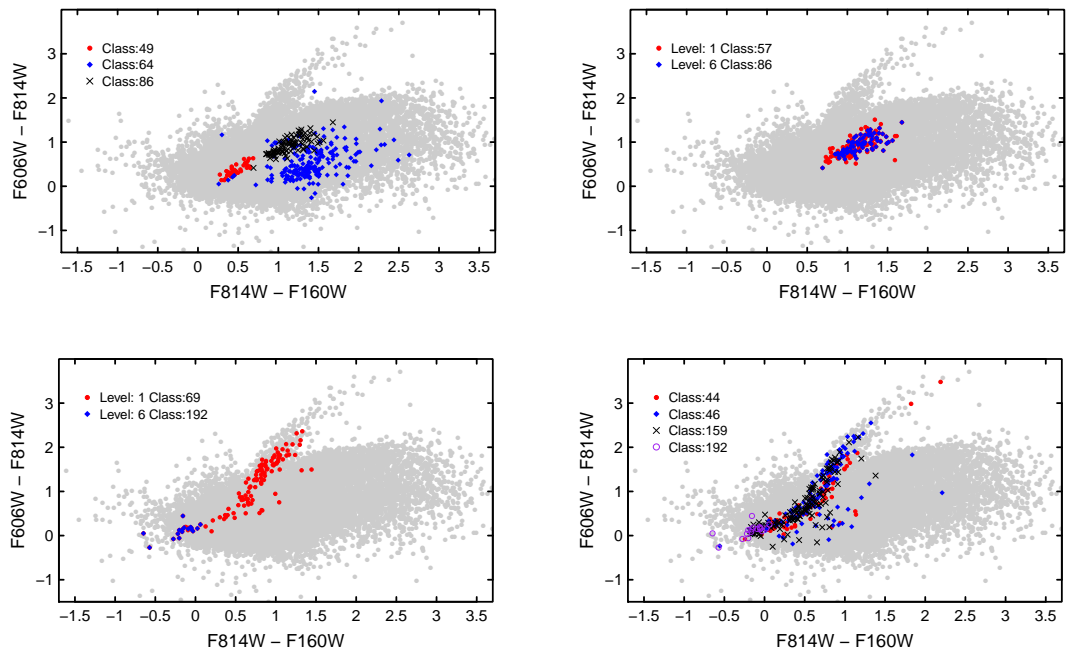

**Figure 9.** These colour-colour diagrams show some of the classification groups in our classified CANDELS catalogue. The background grey points are a random sample of the entire population. In blue, red and black are galaxies from individual classifications. Many of the classifications appear as distinct clusters in colour-colour space. The top right shows galaxies from classification number 57 and one of it's 'child' classifications number 86 which is an example of the hierarchy within the catalogue. The bottom left figure also shows the effect of the hierarchy of classifications, level six being the most detailed classification level, and level one at the highest (coarsest) level. The bottom panels show different classifications for point sources which track the stellar locus; note that in the bottom right panel we find different classifications for sources lying in the same colour space, indicating that, while colour information clearly enters into the classification, the algorithm can offer a more finely controlled object classification and selection.

The first step is to select the pre-set parameters. It was unclear whether the pixel scale and reduced depth compared to the FF images would affect the parameter choices therefore we ran the process multiple times with different options such as two patch sizes (8 and 12 pixels) and two threshold levels ($4\sigma$ and $5\sigma$). On inspection of the results we chose a patch size of 12 pixels and a threshold level of $4\sigma$ above the background level. This produces 9.5 million patches, each of which were then normalised and topologically mapped by GNG to 10,000 GNG nodes. We applied the HC algorithm using the Pearson correlation which resulted in 1,174 groups (using a threshold of 0.045). We select the threshold level based on the quantisation error of the patch groups.

For each of the five fields the connected components step was run to identify galaxies and create the galaxy vector representations that are then grouped together by another HC step using the Pearson correlation. The output is a hierarchy of galaxy classifications. At the top level we choose a minimum number of clusters of 100 and then for each level increment by twenty until the lowest level contains 200 distinct classifications. In addition, we calculate an 'average' galaxy vector representation in each group by averaging the vector representations of all the galaxies in a group. A similarity value between each galaxy and the 'average' galaxy for its group is calculated by computing the Pearson correlation between the vector representations and subtracting it from one. The most similar galaxies will have similarity value of 0. Note that any negative correlations are heavily penalised. This value is important to identify the purest examples in each classification. We provide the similarity scores in the catalogue as 'classification distance' and it is important to use these values to sort the galaxies in each classification.

Choosing the number of clusters is one of the main difficulties of the technique. We have selected a range from 100 to 200 clusters using visual inspection of the classifications to identify which levels create the purest classifications. On inspection of the results the higher granularity of 200 clusters appear to provide the purest classifications. Fortunately the use of hierarchical clustering algorithm makes it straightforward to retrieve different numbers of clusters without requiring significant re-processing.

The catalogue provides classifications for ~60,000 galaxies. Table 1 contains the description of the classification catalogue file. The catalogue file is available in CSV format and we provide a visual version of the catalogue at www.galaxyml.uk. In addition, as each galaxy has a vector representation we can also use the Pearson similarity measure to identify the most similar other galaxies within

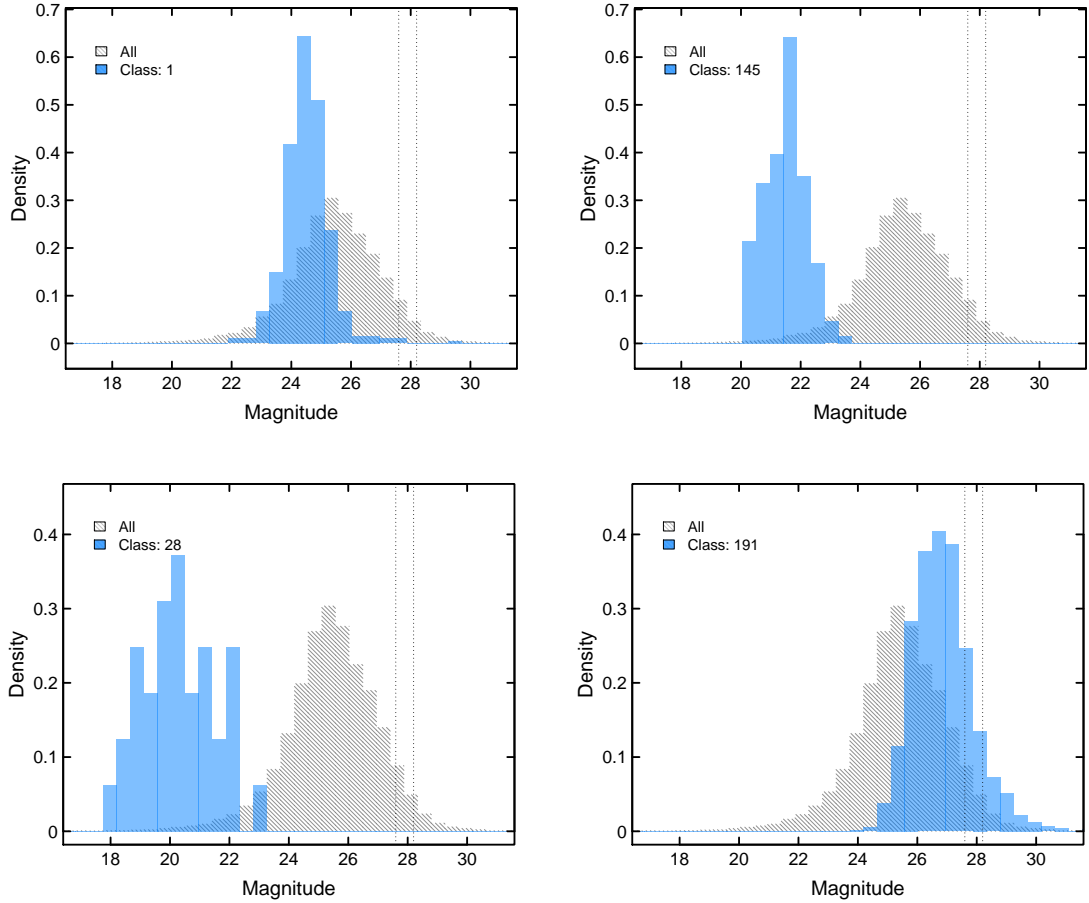

**Figure 10.** These histograms show F606W total magnitudes obtained from the 3D-HST photometric catalogues of Skelton et al. (2014). They compare the magnitude distributions of galaxies given a specific classification (blue) with a random sample of galaxies from the full entire population (grey). The vertical lines are the $5\sigma$ limiting magnitudes for the wide and deep CANDELS surveys. This figure illustrates that the classification process groups galaxies into categories that can be easily described in terms of traditional descriptors such as magnitude, with distinct and 'well behaved' distributions.

CANDELS for each galaxy. We have used this capability to provide a web based galaxy similarity search function at www.galaxyml.uk

To use the catalogue it is important to employ the classification distance column to sort all the galaxies in ascending order. The classification distance columns are shown in Table 1. These distances identify how close each galaxy is to its particular classification. The higher the classification distance the less similar a galaxy is to the classification it is member of.

In order to analyse the classifications and to produce the final catalogue we matched our classification catalogue to the 3D-*HST* catalogues from Skelton et al. (2014), which contain photometry and photometric redshifts for CANDELS. Skelton et al. (2014) determined the photometric redshifts by using EAZY (Brammer et al. 2008).

Figure 9 shows colour-colour plots for some example galaxy classifications, and illustrates the effect of hierarchy: each top level group is split into further levels, which are sub-sets of higher levels. They can be considered increasing levels of detail. Different classifications tend occupy distinct regions of colour space, and it is clear that the stellar locus is clearly delineated. This is not surprising, since colour information is encoded in the classification process (albeit a single colour in this case). Figure 10 shows the F606W total magnitude distributions for selected classifications where again we can see that automatically classified groups tend to have well defined magnitude distributions distinct from the overall population. Finally, photometric redshift distributions are shown in Figure 11; again, showing well defined distributions for different automatic classifications. This demonstrates that the algorithm is actually grouping sources together that can be linked to (or labelled with) well-defined and well-understood observed parameters, and therefore can be put into a practical astrophysical context. Figures 12, 13, 14, 15 are examples of galaxies within different groups and levels within the hierarchical catalogue, illustrating how the algorithm is grouping together similar types of object over a wide dynamic range. While the majority of the classification groups appear well-defined, we note, however, that not all the classification groups are clean. Three examples are shown in Figure 16. They contain inconsistent galaxies, galaxies near the edge of coverage and also galaxies that appear to be outliers.

The catalogue was used to create Figures 12, 14, 15 in the following way. For each field the FITS files for F160W, F814W and F606W were combined into a single PNG image file using

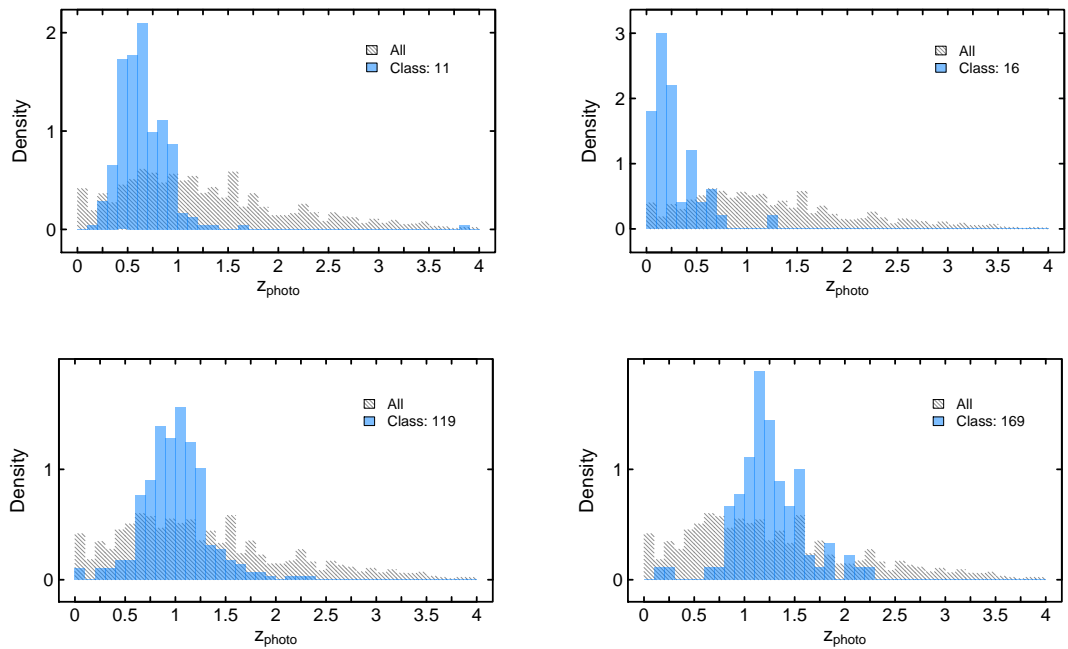

**Figure 11.** We show the photometric redshifts of the galaxies for four different classifications identified by the machine learning technique. The photometric redshifts were obtained from Skelton et al. (2014) who determined them by using EAZY (Brammer et al. 2008). The histogram in grey shows the distribution for a random sample of the full population. As in Figure 10, each classification (based solely on pixel data) falls into well behaved distributions; for example, class 119 (bottom left) clearly contains galaxies at $z \approx 1$. Adding these 'post-processed' labels to automatically classified sources is useful in assigning astrophysical context to the groups the algorithm has identified.

STIFF (Bertin 2012). The catalogue file was then used to identify galaxies in each classification. The galaxies were sorted by their classification distance in ascending order. For example, Figure 15 includes three rows from classification 169. The field, RA and Dec for each galaxy were extracted from the catalogue file. The galaxies were sorted using the classification distance in ascending order. The pixel co-ordinates were identified using the FITS file header and an image thumbnail was cut from the field PNG file around the galaxy. The visual version of the catalogue on the website www.galaxyml.uk showing all 200 classifications for the most detailed classification level was also created using this method.

Figure 13 was produced by manually selecting and ordering ten classifications (23, 174, 6, 86, 45, 8, 11, 140, 30, 146) from the website visual catalogue. These classifications were selected to demonstrate the granularity of classification that is possible using the technique. The catalogue was then used to create the images by repeating the process used for Figures 12, 14, 15.

### 4.1 Identifying Unusual Objects

This technique can be used to identify rarer types of object. An advantage of the technique is that we can use different algorithms in place of hierarchical clustering to achieve a different view of the survey images. One such algorithm is KMeans (Sculley 2010) which can be used in place of hierarchical clustering. We have explored variations of parameters and algorithms in Hocking et al. (2017). We analysed CANDELS with a variant of the machine learning system using KMeans. We scanned the classification groups to identify which groups contained galaxies with large elliptical central bulges but with localised higher emission in the shorter wavelengths - this could be the result of mergers or conjunctions with background galaxies. We identified a strong lensing galaxy that is currently known in the NASA Extragalactic Database (NED) and we found two candidates as seen in Fig 17 which are not classed as lenses in NED.

**Table 1.** The format and columns of the catalogue produced by the machine learning technique.

| Column Position | Column Name | Description |
|---|---|---|
| 1 | Field Id | The identifier of the field where the object resides. 0 GOODS-N, 1 UDS, 2 EGS, 3 COSMOS, 4 GOODS-S |
| 2 | Object Id | The ID of the object from the 3D *HST* catalogue by Skelton (based on a cross match) |
| 3 | RA (degrees) | Right Ascension (J2000) |
| 4 | Dec (degrees) | Declinaton (J2000) |
| 5-10 | Classifications | Hierarchical classifications, 6 levels of classifications |
| 11-16 | Classification distances | A number between 0 and 1. The nearer to 0 the more relevant the galaxy is to the classification. These fields are important for sorting objects within classifications. |

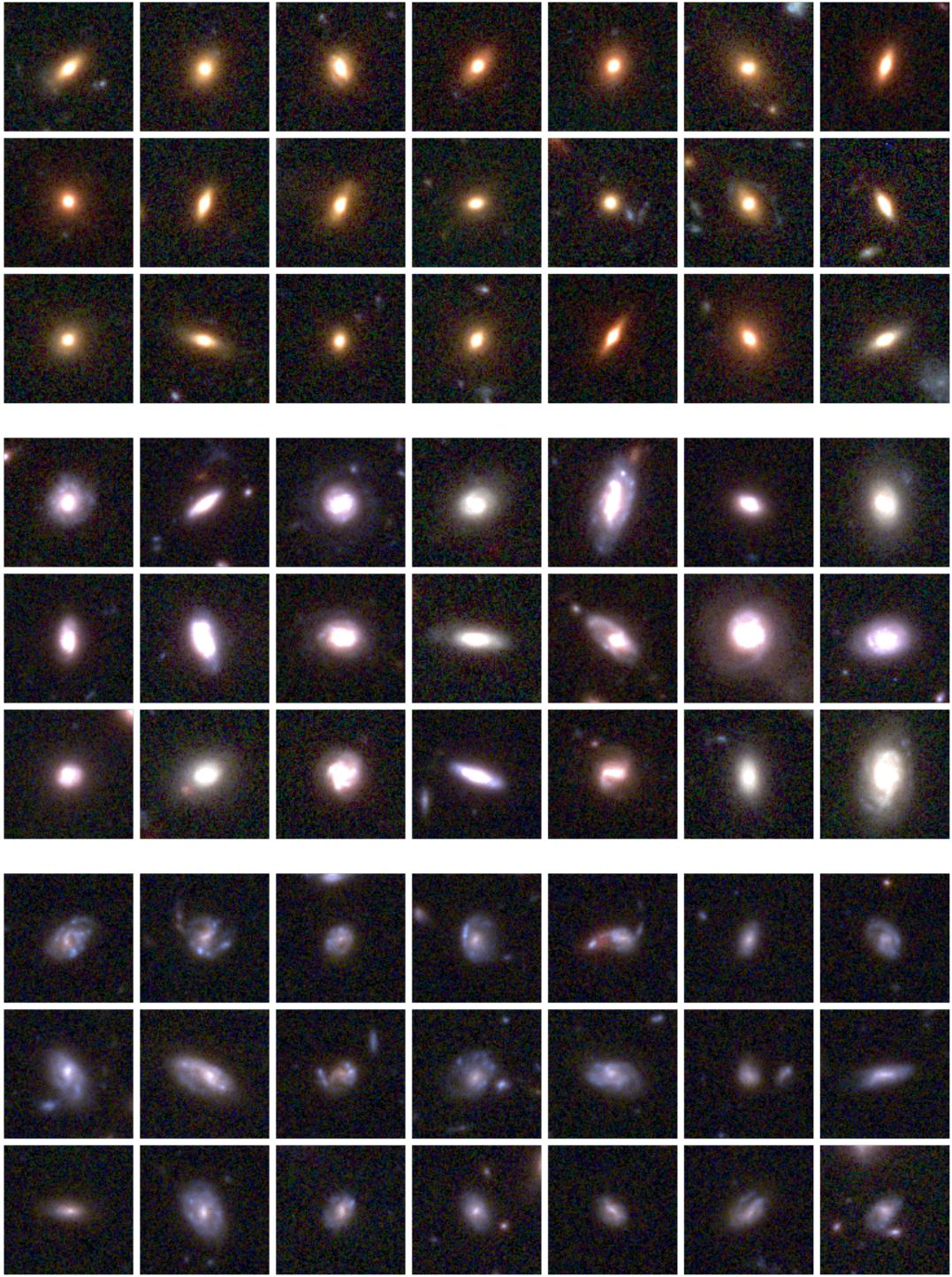

**Figure 12.** Example images from the top level of three different CANDELS classification groups (classification groups 7, 18 and 98). Each image is $6'' \times 6''$. The galaxies in each group are ordered row-wise in order of their similarity to the 'average' classification in the parameter space of the group. The top left image is the most similar galaxy to the 'average' and the bottom right is most dissimilar. The classification catalogue provides these as classification distances which can proxy as a quality flag. The distances are normalised between 0 and 1, with 0 being an identical match to the average. Here the RGB channels are the F160W, F814W and F606W bands, but note that the latter was not included in the learning.

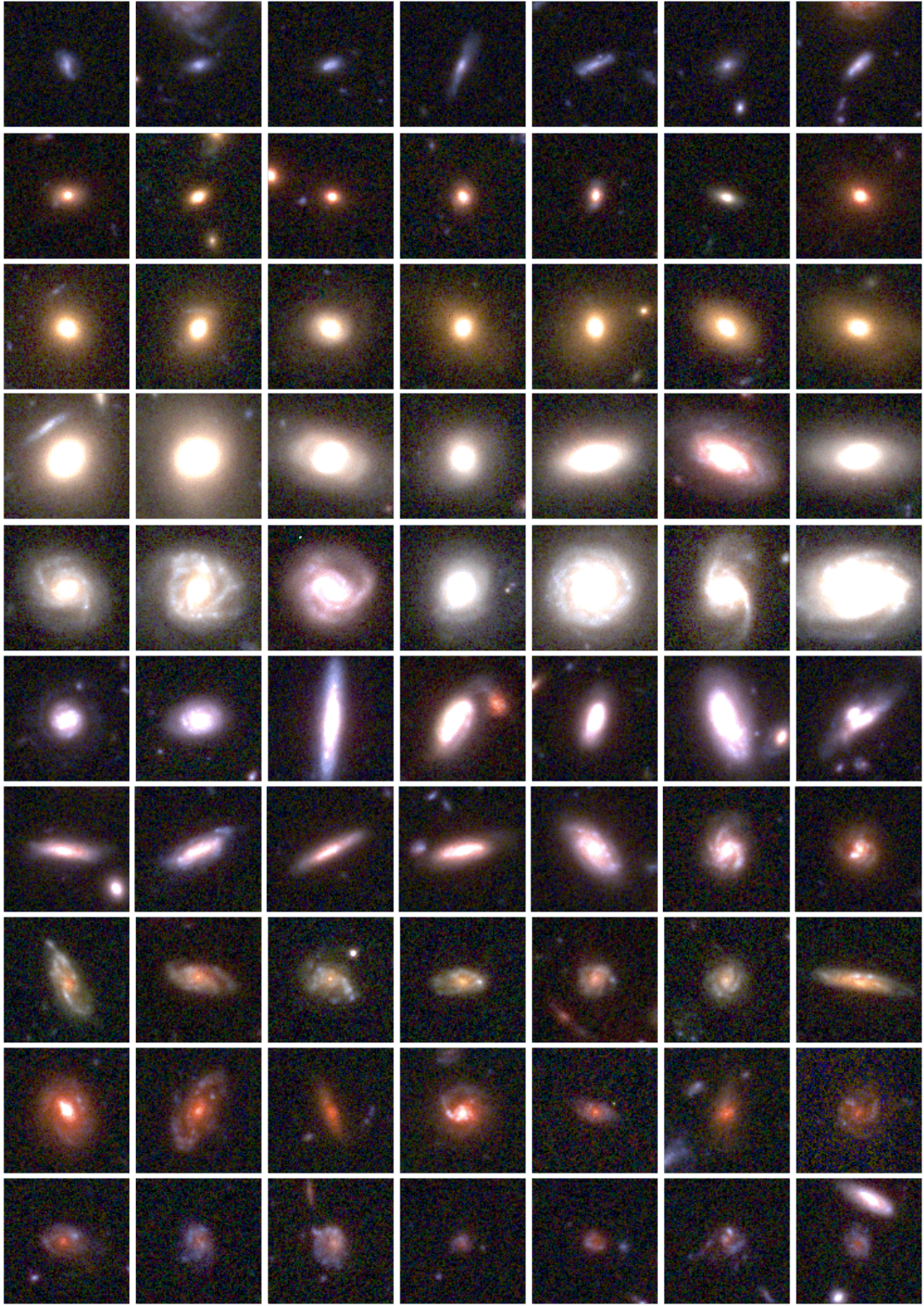

**Figure 13.** Each row of $6'' \times 6''$ images shows galaxies in an individual classification group, and are selected from the lowest hierarchy level in that group. The galaxies are ordered left to right by their similarity to the average galaxy with the first panel most similar to the average. Again, the similarity of sources in each group is clear. The RGB channels are the F160W, F814W and F606W bands.

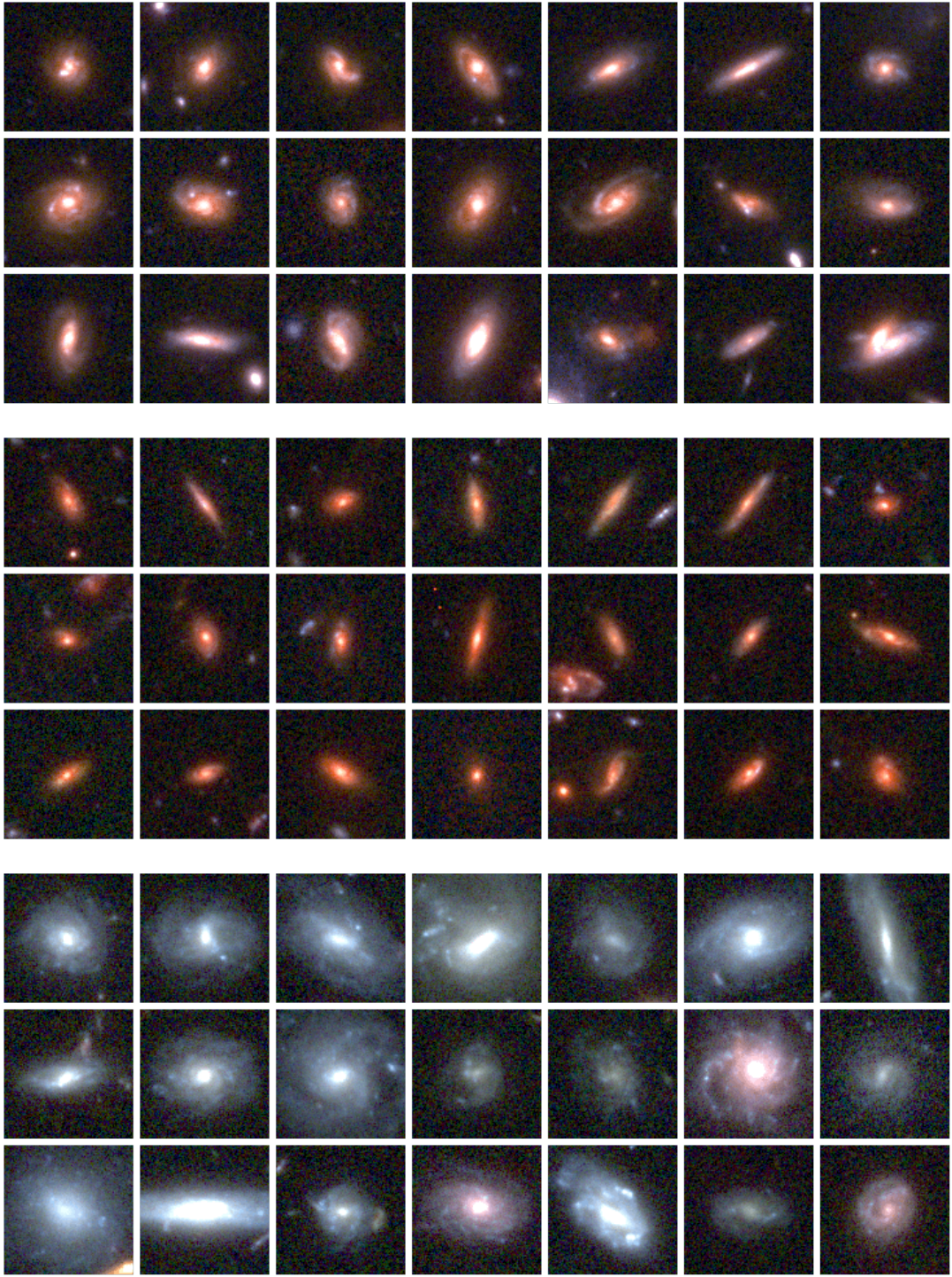

**Figure 14.** Examples of galaxies in three classification groups (30, 36, 48) from level one (the coarsest classification) in the hierarchy. As before, each image is $6'' \times 6''$ and ordered left to right in order of similarity to the 'average' galaxy in the group. The RGB channels are the F160W, F814W and F606W bands.

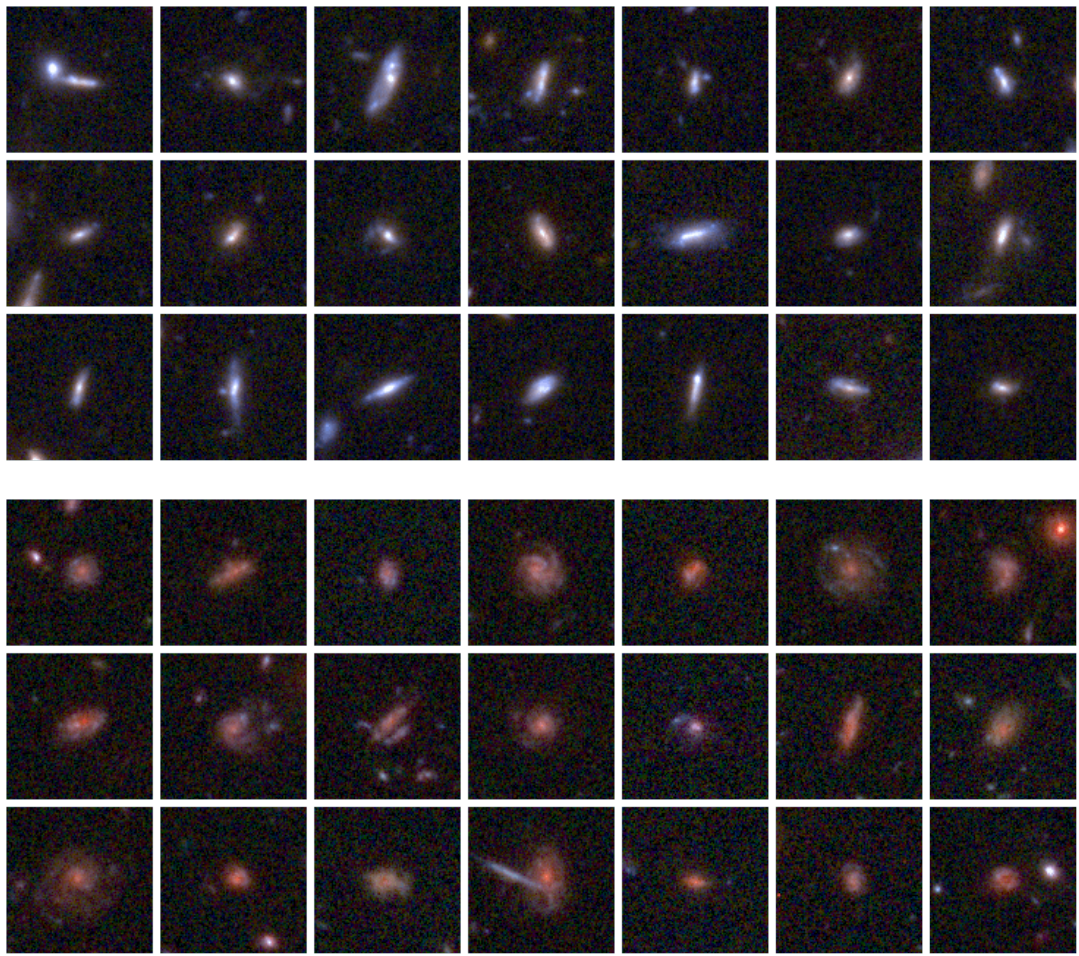

**Figure 15.** Examples of galaxies in two classification groups: 8 at level one (low level of refinement) and 169 at level six (higher level of refinement). The images are $6''\times 6''$ and the RGB channels are the F160W, F814W and F606W bands.

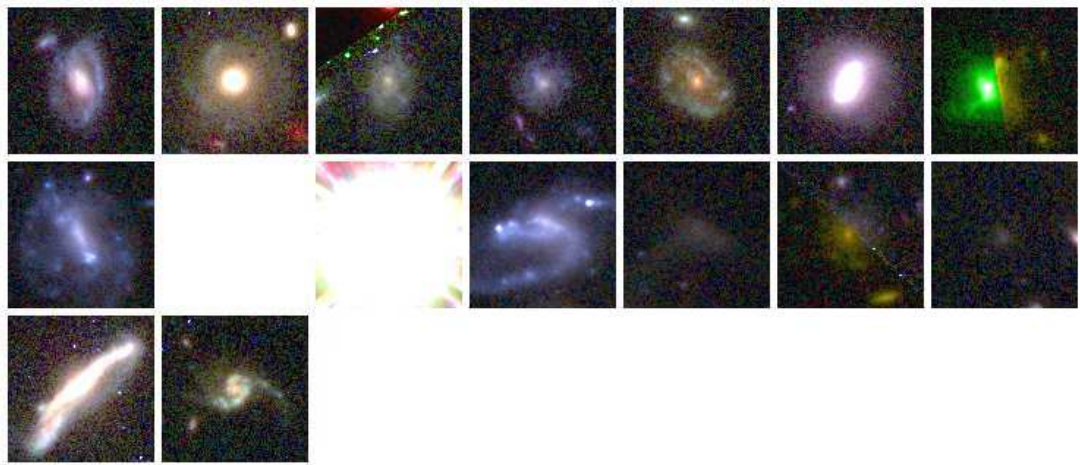

**Figure 16.** The majority of the classifications groups are very clean. However, there are some that are less so such as these three classifications: 24, 41 and 56. Each row is an individual classification. The third row appears to include objects that are outliers distinct from other galaxy classifications.

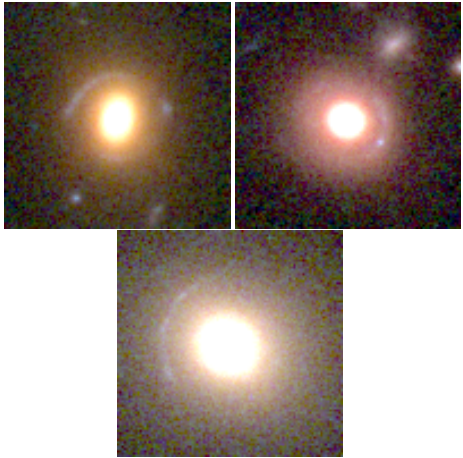

**Figure 17.** Two potential strong lensing candidates (top) and a known lens (bottom). All three appear in the same classification group. The galaxy to the top left is in UDS at location $02^h17^m06^s.20$ and $-05°13'17''.6$. The galaxy to the top right is in EGS at $14^h19^m00^s.12$ and $+52°42'48''.9$.

### 4.2 Comparison to the Galaxy Zoo CANDELS Project

Galaxy Zoo (GZ) has been providing crowd-sourced statistically robust visual morphological classifications for some years now (Lintott et al. 2008, 2011; Willett et al. 2013). They have turned their attention to CANDELS and have recently published detailed morphological classifications for three of the CANDELS fields: GOODS-South, UDS and COSMOS. The classifications were provided by 95,000 volunteers with each galaxy receiving an average of 43 classifications (Simmons et al. 2016). GZ leads a volunteer agent through a decision tree of questions about an individual galaxy. Depending on the answer to a question the agent will follow different paths down the decision tree. Between two and nine questions are asked of the volunteer; for example, if the galaxy image is a star or artifact then only two questions are required, if it is a spiral galaxy then up to nine questions are required. These classifications have been consolidated and robustly analysed by the GZ team to provide a catalogue of weighted fractions for each answer in the tree for 48,000 galaxies. Simmons et al. (2016) describe the catalogue, the methodology and provide a detailed analysis.

How do the machine learnt classifications and the human classifications compare? Clearly we cannot expect a direct mapping between GZ classifications and our hierarchical grouping, but we can use the GZ catalogue to ask the question of whether our groupings would have had a 'concordance' classification (based on the questioning tree) from a cohort of human classifiers. GZ provides two catalogue files. One is the full catalogue for COSMOS, UDS and GOODS-South and a second that contains adjustments to the classifications made to the deep survey for GOODS-South. We choose to compare our data with the original catalogue as no adjustment has been made to the machine learning technique for the different depths. The catalogue files provided by GZ include the number of classifications and the weighted and unweighted fractions for each answer in the decision tree. We consider three top-level questions: T00 A0 'is the target smooth and rounded?', T00 A1 'does it contain features or a disk?' and T00 A2 'is it a star or artifact?'. The weighted fractions run from 0 to 1 corresponding to a negative or affirmative result. We ask whether the algorithm has assembled groups for which ≥50% of the members (that have GZ classifications) have answers to any of these questions above a weighted fraction of 0.5. We call this a 'concordance' classification. We can find several examples of concordance classifications, and we show two examples of each in Figures 16, 17 and 18, presenting the top seven galaxies from each classification as a guide, and the distributions of the weighted fractions of the answers to each of T00 A0–2 for galaxies in each group.

GZ also includes several 'clean' classifications, where a boolean flag is assigned to a subset of GZ classifications for which the weighted classification indicated a high conviction for 'clean_feature' (229), 'clean_spiral' (278), 'clean_smooth' (4662), 'clean_edge_on' (162) and 'clean_clumpy' (215). The numbers in parentheses indicate the number of clean classifications in the matched catalogue. Note that the majority are 'clean_smooth'. The top level of our hierarchical classification contains 100 groups, and can be considered the coarsest level of classification refinement. This is probably most suitable for this comparison: we can simply assess the fraction of machine learnt groups that contain each of the GZ clean classifications. One could argue that if a high fraction of clean classifications are contained within a small fraction of top level machine learnt groups, then the algorithm has successfully pigeon-holed the human classifications. On the other hand, these clean descriptors are rather broad, whereas even the coarsest level of machine learnt classification offers a way to segregate (for example) 'smooth' galaxies.

We consider each of the clean classifications described above and sort the list of top level machine learnt groups according to the number of galaxies matched to the clean lists. We then simply calculate the cumulative fraction of each clean list to assess the fraction of unique groups containing 50% and 100% of the clean classification galaxies. The results are given in Table 2, which lists the 50% and 100% fractions describing how the various clean classes are distributed within our machine learnt groups. For spiral, featured, clumpy and edge-on galaxies, the majority of the cleanly classified galaxies are contained within less than 10% of the top level groups. The fraction is slightly higher for the smooth class. In all but the smooth class, 100% of the clean classifications are contained within around 50% of the machine learnt groups. For the smooth classification this is much higher – the galaxies seem to be spread over the majority of the machine learnt groups. This is perhaps unsurprising because the smooth classification dominates the clean class galaxies, and our algorithm has segregated these into a diverse set of sub-classes even at the top level of our hierarchical classification. Still, the fact that in all cases around half of the clean classifications are described by a minority of machine classes suggests that the algorithm is automatically classifying targets in a manner that is not dissimilar to a human inspector.

We conclude this section with a suggestion of an additional

**Table 2.** Fraction of top level machine learnt (ML) classification groups containing 50% and 100% of the galaxies in the various Galaxy Zoo 'clean' classes.

| Galaxy Zoo clean class | Fraction of ML groups containing 50% of clean class | Fraction of ML groups containing 100% of clean class |
|---|---|---|
| Smooth  | 13% | 89% |
| Spiral  | 7%  | 52% |
| Featured | 5% | 53% |
| Clumpy  | 8%  | 51% |
| Edge-on | 5%  | 47% |

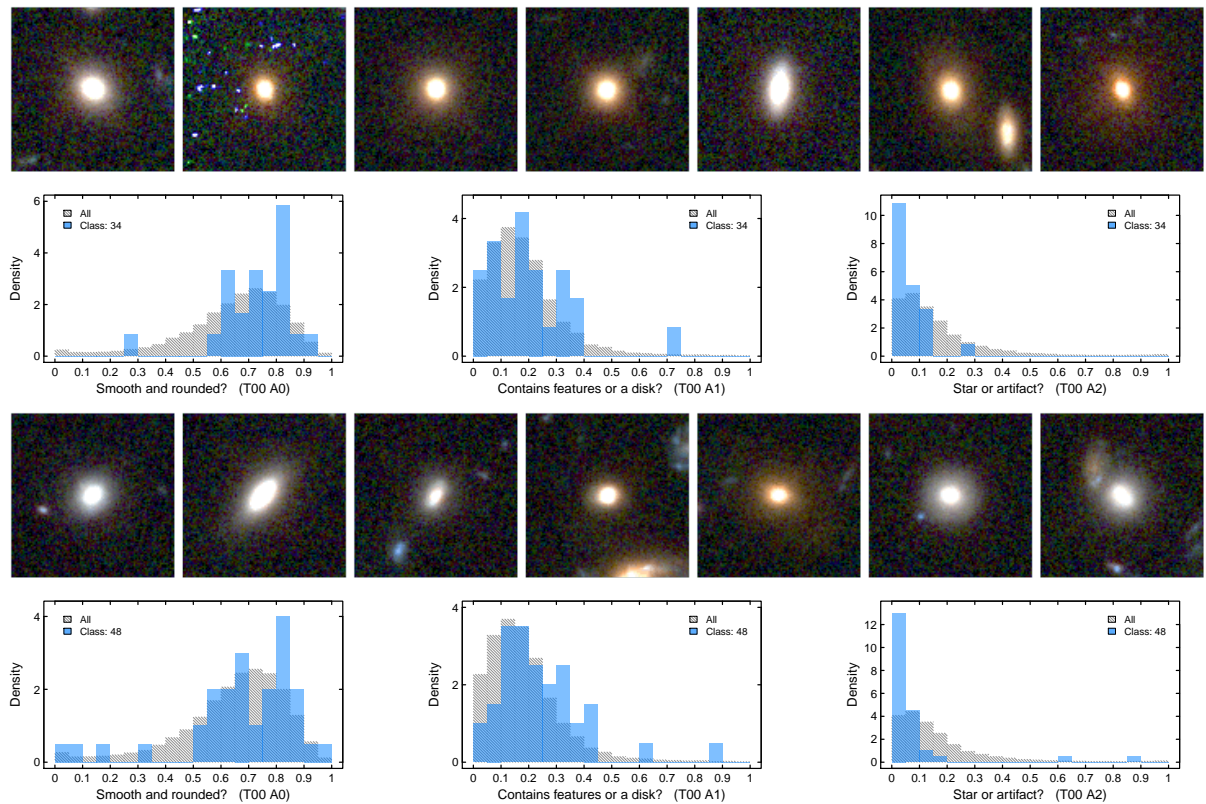

**Figure 18.** This figure shows two examples of what we refer to as a 'concordance' group, where over 50% of the galaxies for which Galaxy Zoo classifications were made have a weighted fraction over 0.5 for question T00 A0 'is the target smooth and rounded?'. The images show the top seven matches in the group and the histograms compare the distribution of weighted fractions for questions T00 A0, A1 and A2 (see text) for galaxies in the group (blue histogram) compared to the full range of GZ classifications (grey histograms). Although not every machine learnt grouping can be described as a concordance group when compared to Galaxy Zoo classifications, this figure illustrates that the algorithm is creating groups that would have received a consistent human classification.

## 5 SUMMARY

We present an efficient unsupervised machine learning algorithm that uses a combination of growing neural gas, hierarchical clustering and connected component labelling to explore surveys by automatically segmenting and labelling imaging data. The technique is a patch based model that doesn't process whole images of galaxies. Instead, it processes many small overlapping patches from survey images. Each small overlapping patch is typically much smaller than the size of a galaxy such as a section of a spiral arm, or a section of a low surface brightness feature. The combination of a patch based model and graph algorithm is a novel technique previously unseen in astronomy. In addition, unlike existing unsupervised techniques, the features of the technique are very simple and as such are very computationally efficient enabling it to process potential use for this technique which is to make predictions on which galaxies will be classified as, for example, clean_spiral by human classifiers. Indeed, blending the machine learning and human classification methods might be a particularly powerful technique; for instance, for extremely large samples of galaxies (or just large images), the algorithm could perform a 'first pass' unsupervised classification and feed subsamples of those results (blindly) to a cohort of human inspectors.

FITS surveys, whereas existing techniques typically convert FITS survey images into normalised JPG image stamps of individual galaxies.

As a demonstration we have applied the algorithm to images from the *HST* Frontier Fields survey, showing how the algorithm can examine data from one field (Abell 2744) and search another 'unseen' image (MACS 0416.1−2403) to successfully classify galaxies that would be classified as 'early' and 'late' types by a human inspector. From this trivial example we apply the algorithm to all five *HST* CANDELS fields, producing a catalogue of ∼60,000 galaxy automatic classifications. The catalogue, visual catalogue and galaxy similarity search is available at www.galaxyml.uk. We demonstrate how the automatic classifications have distinct distributions in more familiar parameter spaces such as magnitude, colour and redshift, allowing for post-labelling to place them in an astrophysical context ($z \approx 1$ red spiral, etc.). By comparison to crowd sourced classifications for thousands of the same galaxies in the Galaxy Zoo project, we also demonstrate that many of our automatic groupings have a 'consensus' classification from a large cohort of human inspectors.

One simple way of utilizing the CANDELS classification catalogue is to use it to assemble samples of galaxies (or stars) that are photometrically and morphologically similar to a given test example. For example, one might have detailed observations of a specific galaxy in CANDELS and desire to find more examples of similar objects to build a statistical sample. One could simply match this target to the classification catalogue to find out which classification group it resides in, and therefore find all the other galaxies that 'look' (as far as the feature space allows) similar to it. Naturally, the selection function for this exercise would be complicated to understand (i.e. difficult to express in terms of, say, colour cuts), and that might be a limitation of this approach.

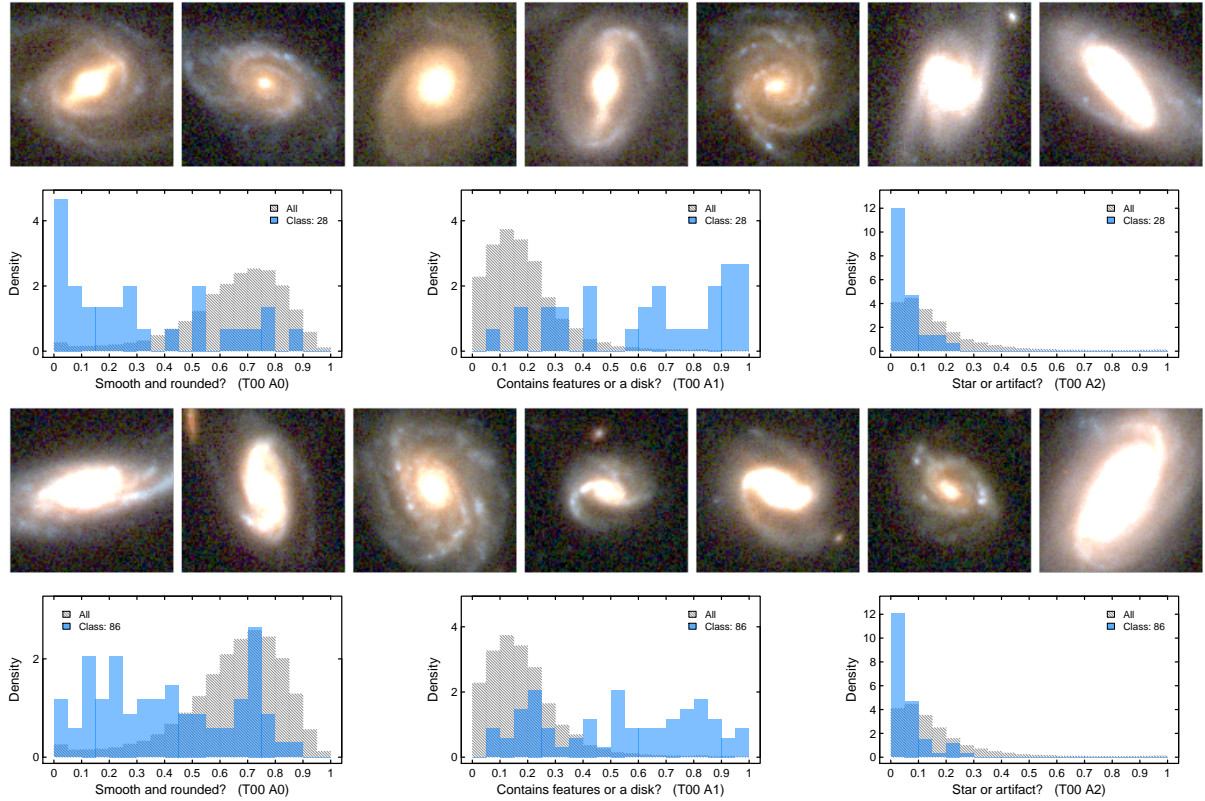

**Figure 19.** This figure shows two examples of what we refer to as a 'concordance' group, where over 50% of the galaxies for which Galaxy Zoo classifications were made have a weighted fraction over 0.5 for question T00 A1 'does it contain features or a disk?'. The images show the top seven matches in the group and the histograms compare the distribution of weighted fractions for questions T00 A0, A1 and A2 (see text) for galaxies in the group (blue histogram) compared to the full range of GZ classifications (grey histograms). Although not every machine learnt grouping can be described as a concordance group when compared to Galaxy Zoo classifications, this figure illustrates that the algorithm is creating groups that would have received a consistent human classification.

The unsupervised nature of the algorithm allows for the discovery of features not previously known; this will be useful for data discovery in the era of extremely large surveys such as *Euclid* and LSST. The feature space that is mapped by the algorithm is completely arbitrary, and could involve a large number of parameters not explored here (where we have concentrated on pixel intensity distribution). For example, in the case of LSST, one could introduce the time domain into the classification process, affording the ability to automatically identify and classify transient phenomena.

There are limitations to the method that should be noted. The most significant is the choice of the initial data matrix. In this work we use sample vectors that effectively encode information about colour and intensity distribution on small (few pixel) scales. In principle the sample vector can be arbitrarily large, but at the cost of computation time; therefore there is a balance between performance and the sophistication of the data matrix. It is clear that the exact choice of data matrix will have an impact on the ability of the algorithm to successfully segment and classify input data. It is possible that one could use an algorithm that identifies the optimal set of features to use (see unsupervised feature learning in Bengio et al. (2013), also stacked denoising autoencoders by Vincent et al. (2010) ), but that is beyond the scope of the current paper.

We have not fully optimized the algorithm for speed (and as noted above, performance will depend on the complexity of the data matrix), however as a guide, in the example presented here the training process on the Abell 2744 imaging took 36 msec per pixel and the application of the trained algorithm to the new MACS 0416.1−2403 image took 1.5 msec per pixel. The work was performed on a desktop Intel Core i7-3770T 2.50GHz with 8GB RAM. These performances can clearly be dramatically improved, especially through the use of GPUs and optimal threading. The classification process is fully parallelisable, and the compute time for classification scales linearly with the number of pixels for a given model, making this a highly efficient algorithm to apply to large imaging data.

We conclude by noting that the algorithm presented here is not limited to imaging data: spectral data could also be passed through the process, which may be relevant to current and next generation radio surveys. Indeed, the algorithm is completely general and one can envision applications beyond astronomy, for example in medical or satellite imaging.

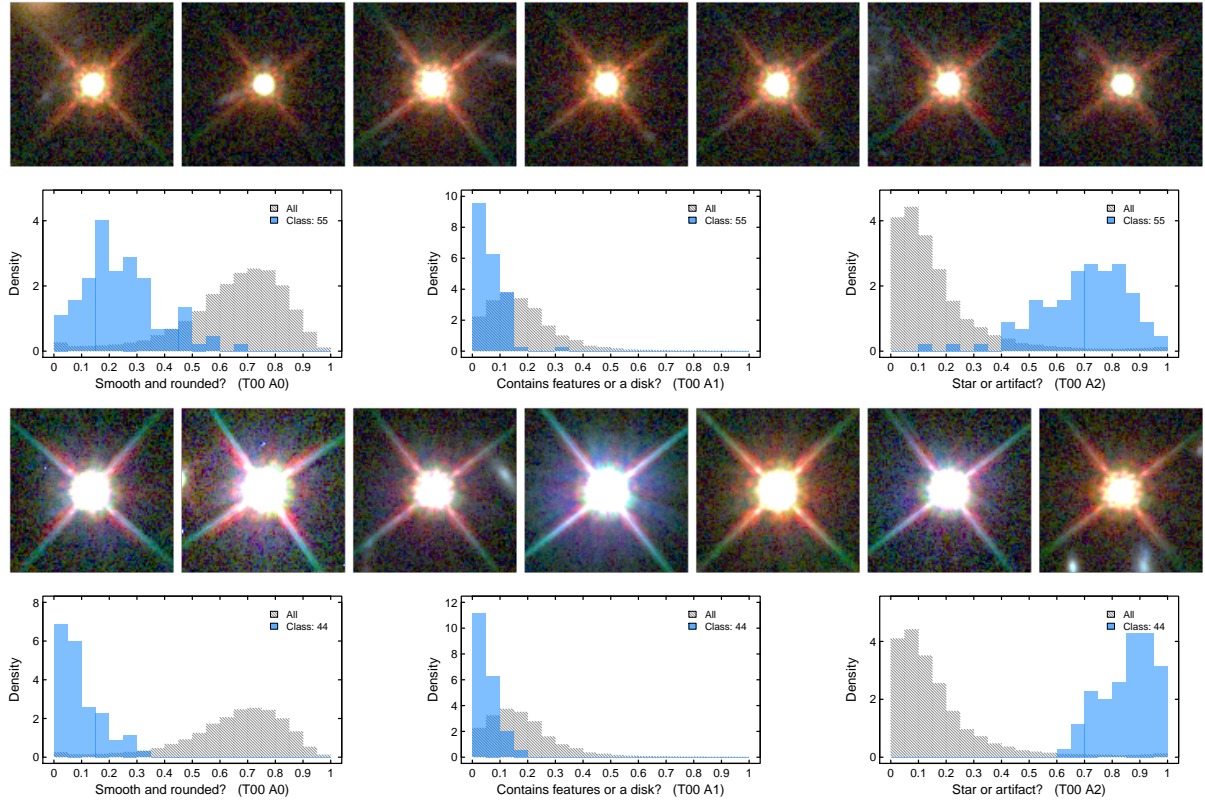

**Figure 20.** This figure shows two examples of what we refer to as a 'concordance' group, where over 50% of the galaxies for which Galaxy Zoo classifications were made have a weighted fraction over 0.5 for question T00 A2 'is the target a star or artifact?'. The images show the top seven matches in the group and the histograms compare the distribution of weighted fractions for questions T00 A0, A1 and A2 (see text) for galaxies in the group (blue histogram) compared to the full range of GZ classifications (grey histograms). Although not every machine learnt grouping can be described as a concordance group when compared to Galaxy Zoo classifications, this figure illustrates that the algorithm is creating groups that would have received a consistent human classification.

## ACKNOWLEDGEMENTS

The authors thank N. Hine for comments on the manuscript. J.E.G. acknowledges the support of the Royal Society through a University Research Fellowship. We gratefully acknowledge the support of NVIDIA Corporation with the donation of the Tesla K40 GPU used for this research.

This paper has been typeset from a T<sub>E</sub>X/L<sup>A</sup>T<sub>E</sub>X file prepared by the author.